\documentclass[aps,pra,showpacs,twocolumn,longbibliography]{revtex4-1}

\usepackage{color}
\usepackage[utf8]{inputenc}
\usepackage[english]{babel}
\usepackage{graphicx}
\usepackage{hyperref}
\usepackage{epstopdf}
\usepackage{amsfonts}
\usepackage{amsmath}

\newcommand{\ie}{i.\,e.\ }

\newcommand{\cf}{cf.\ }

\newcommand{\fref}[1]{\text{Fig.}~\ref{#1}}
\newcommand{\eref}[1]{\text{Eq.}~\eqref{#1}}

\begin{document}

\title{Generating a stationary infinite range tractor force via a multimode optical fiber}
\author{C. A. Ebongue}
\author{D. Holzmann}
\author{S. Ostermann}
\author{H. Ritsch}
\email{helmut.ritsch@uibk.ac.at}
\affiliation{Institut f\"ur Theoretische Physik, Universit\"at Innsbruck, Technikerstraße 21, A-6020 Innsbruck, Austria}

\begin{abstract}
Optical fibers confine and guide light almost unattenuated and thus convey light forces to polarizable nano-particles over very long distances. Radiation pressure forces arise from scattering of guided photons into free space while gradient forces are based on coherent scattering between different fiber modes or propagation directions. Interestingly, even scattering between co-propagating modes induces longitudinal forces as the transverse confinement of the light modes creates mode dependent longitudinal wave-vectors and photon momenta. We generalize a proven scattering matrix based approach  to calculate single as well as inter-particle forces to include several forward and backward propagating modes. We show that an injection of the higher order mode only in a two mode fiber will induce a stationary tractor force against the injection direction, when the mode coupling to the lower order mode dominates against backscattering and free space losses. Generically this arises for non-absorbing particles at the center of a waveguide. The model also gives improved predictions for inter-particle forces in evanescent nanofiber fields as experimentally observed recently. Surprisingly strong tractor forces can also act on whole optically bound arrays.
\end{abstract}


\maketitle 

\section{Introduction}
In an optical fiber, light is transversely confined and  transmitted over very long distances without attenuation. The field can be well decomposed in transverse modes, each of them associated with a specific transverse field pattern and a corresponding longitudinal propagation wave vector. For specially designed fibers as optical nano-fibers or hollow core fibers a significant fraction of the light intensity is propagating in free space outside the actual fiber material and can interact with particles in this region~\cite{vetsch2010optical, marksteiner1994coherent}.

If one places a nanoscopic particle or even a single atom into the fiber fields, it will significantly perturb the light propagation and redistribute the field among different modes and propagation directions. As the light carries momentum, the corresponding photon redistribution leads to a net optical force\cite{maimaiti2016nonlinear}. The magnitude and direction of this force strongly depends on the fiber and particle geometry and the properties of the injected field. For several particles coupled to the same fiber, collective scattering enhances these forces and creates strong inter-particle forces depending on their relative distance. This leads to optical binding, selfordering and nonlinear motional dynamics~\cite{asboth2008optomechanical,chang2013self,griesser2013light}. 

In some recent work we exhibited that important properties of these forces can be well understood from a generalized beam splitter approach involving only two fiber modes and an additional free space loss mode~\cite{maimaiti2016nonlinear}. Surprisingly this model also allows to identify a parameter regime, where the total force on a particle points against the wave-vector of the impinging light, i.e. we get a tractor beam force~\cite{brzobohaty2013experimental}. This feature even persists for arrays of several bound particles.

Physically the tractor beam phenomenon is tight to the fact, that scattering from a higher order to a lower order mode requires a momentum contribution from the particle along the field propagation. Thus due to momentum conservation the particle is thus pushed in the opposite direction. The momentum exchange is, however, smaller than for backscattering a photon into the opposite direction. Hence for a more realistic calculation one needs to  include the backscattered fields and absorption as competing processes. 

Note that  injecting phase coherent superpositions of light simultaneously into two transverse modes can create 3D optical trapping positions along the fiber~\cite{le2004atom}. Particles in these traps can be moved along the fiber in any direction by moving the traps via a relative phase control of the two modes in time~\cite{sadgrove2016quantum}. This is different from our tractor beam mechanism, which is independent of position along the fiber and does not require external  phase control.
   
Here we will investigate the properties and limitations of implementing such a generic tractor beam mechanism for small beads trapped in an optical two-mode nanofiber. While a full numerical calculation using 3D finite element software is possible~\cite{maimaiti2015higher}, it does not allow to scan large parameter and size ranges. Hence it is difficult to get good qualitative understanding and the whole range of possibilities offered by such a system.

 Hence to better get the central idea, we will first study a rather idealized two mode model including only for forward scattering and some general absorption losses to identify the key parameter region for the appearance of tractor forces. In a second step we will generalize to a more realistic description including the backscattered fields and loss to free space. While it seems difficult to find realistic favorable parameters for a tractor forces in multi mode nanofibers, hollow core fibers seem much more promising. Finally we go beyond the case of single particles and study how ordering and optical binding can be combined with tractor forces. 
\begin{figure}
\centering
\includegraphics[width=0.45\textwidth]{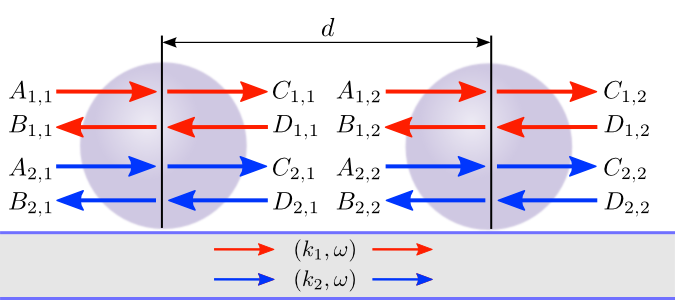}
\caption{(Colour online)  Schematic picture of two micro-spheres trapped close to a optical two mode fiber. In the presented model the spheres are approximated as beam splitters and the fields are determined by the incoming and outgoing amplitudes. The red arrows denote the amplitudes for the first (fundamental) mode and blue arrows characterize the second (higher order) mode. The first index for the amplitudes defines the mode number whereas the second index stands for the particle number. An analogous picture holds for a particle in the field of a hollow waveguide~\cite{domokos2002quantum,gomez2001resonant}.}
\label{fig:scheme}
 \end{figure}

\section{Model}
We consider $N$ polarizable spherical particles within the field of a multimode optical waveguide~\cite{domokos2002quantum,gomez2001resonant,horak2003giant}. Such a geometry can either be realized by placing the particles inside a multi-mode hollow core fiber~\cite{russell2003photonic,renn1995laser} or trap them close to the surface of a tapered nanofiber with strong evanescent field components~\cite{maimaiti2016nonlinear}. The beads in the field are modeled as effective beam-splitters as shown in fig.~\fref{fig:scheme}, which couple the local field mode amplitudes left and right of the beam splitter. These mode amplitudes are connected via an effective scattering matrix representing the underlying microscopic scattering processes integrated over the bead volume. As the bead only interacts with a spatial fraction of the mode, this matrix on the one hand includes forward and backward scattering of photons into the same transverse fiber mode as well as on the other hand it allows cross-coupling between the fundamental and the higher order transverse modes. 

\begin{figure}
\begin{minipage}{0.49\textwidth}
\centering
\includegraphics[width=0.6\textwidth]{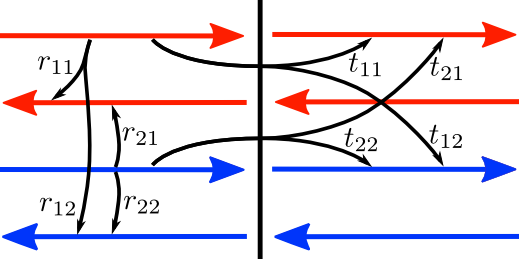}
\caption{(Colour online) The fundamental processes which are described by the presented model. All reflection processes are shown on the left side of the beam-splitter and all transmission processes are shown towards the right side of the beam-splitter. Of course, all processes are mirror symmetric and can also occur into the opposite direction.}
\label{fig:processes}
\end{minipage}
 \end{figure}

As depicted in~\fref{fig:processes} the amplitudes $0\le t_{ij}\le 1$ describe forward scattering processes (into the same mode for $i=j$ and into any other mode for $i\neq j$) the reflection coefficients $0\le r_{ij}\le 1$ describe reflections into the same mode as well as mode mixing reflections into the other transverse mode (\cf~\fref{fig:processes}). In principle all these coefficients have to be derived from solving the corresponding Helmholtz equations for any specific implementation and boundary conditions. This has to be done numerically or by fitting experimental results~\cite{maimaiti2016nonlinear}. We will  stay with a general approach for the moment and only give some estimates for specific examples in Appendix A.   

At this point we thus end up with four independent mode amplitudes connected via a four-port beam splitter matrix.  The most general matrix, which describes the occurring processes as they are depicted in~\fref{fig:processes} takes the form
\begin{equation}
\textbf{M}_{4p}=\begin{pmatrix}
   t_{11}e^{i\phi_{11}} & t_{21}e^{i\phi_{21}}& r_{11}e^{i\psi_{11}}& r_{21}e^{i\psi_{21}}\\
   t_{12}e^{i\phi_{12}} & t_{22}e^{i\phi_{22}}& r_{12}e^{i\psi_{12}}& r_{22}e^{i\psi_{22}}\\
   r_{11}e^{i\psi_{11}} & r_{21}e^{i\psi_{21}}& t_{11}e^{i\phi_{11}}& t_{21}e^{i\phi_{21}}\\
   r_{12}e^{i\psi_{12}} & r_{22}e^{i\psi_{22}}& t_{12}e^{i\phi_{12}}& t_{22}e^{i\phi_{22}} \end{pmatrix}.
\label{eqn:M4p}
\end{equation}
It connects the four input and output fields of the $j$th particle via
\begin{equation}
  \begin{pmatrix} C_{1,j}\\C_{2,j}\\B_{1,j}\\B_{1,j}\end{pmatrix}=\textbf{M}_{4p}\begin{pmatrix}
  A_{1,j}\\A_{2,j}\\D_{1,j}\\D_{2,j}\end{pmatrix}.
\label{eqn:conn4p}
\end{equation}
Due to the fact that photon losses are neglected within our treatment the matrix~\eqref{eqn:M4p} has to fulfil the unitary condition $\textbf{M}_{4p}^\dagger \textbf{M}_{4p}=\mathbb{I}$. This ensures that the number of photons is conserved. A generalization to include some internal loss or scattering to other modes is straightforward and certainly helps to quantitatively model a specific experiment, but will not give qualitatively new insights. 

In order to make statements about a tractor beam behavior of the injected light fields, we need to calculate the light induced force onto the particles. This can be done by following proven recipes~\cite{asboth2008optomechanical,maimaiti2016nonlinear}. We calculate the forces acting on the particle via a Maxwell stress tensor based approach. In the effective 1D geometry of a fiber, it suffices to consider the  Minkwoski's  photon momenta $P$ inside a medium with refractive index $n$
\begin{equation}
P=n\hbar k.
\end{equation}
The total momentum of the light propagating in the $i$-th ($i\in\{1,2\}$) mode at the left and right side of the $j$-th bead can be expressed in terms of the propagating photon numbers
\begin{align}
N^\mathrm{L}_{i,j} &= \frac{cn\epsilon_0}{2}\left(|A_{i,j}|^2+|B_{i,j}|^2\right),\\
N_{i,j}^\mathrm{R} &= \frac{cn\epsilon_0}{2}\left(|C_{i,j}|^2+|D_{i,j}|^2\right)
\end{align}
at each side of the bead. The momentum left and right of the beads then reads
\begin{align}
P_{tot}^\mathrm{L}=N_{1,j}^\mathrm{L}\hbar k_1 + N_{2,j}^L\hbar k_2,\\   
P_{tot}^\mathrm{R} = N^\mathrm{R}_{1,j}\hbar k_1 + N^\mathrm{R}_{2,j}\hbar k_2 .
\end{align}
The force on the $j$th bead thus can be simply found from the missing momentum which results in
\begin{equation}
F_j=\hbar k_1(N_{1,j}^L-N^R_{1,j})+\hbar k_2(N_{2,j}^\mathrm{L}-N_{2,j}^\mathrm{R}). \end{equation}
Hence, the force onto the $j$th beam-splitter in terms of the amplitudes is given as
\begin{align}
F_j=\frac{cn\epsilon_0\hbar}{2}(&k_1(|A_{1,j}|^2+|B_{1,j}|^2-|C_{1,j}|^2)\\
+&k_2(|A_{2,j}|^2+|B_{2,j}|^2-|C_{2,j}|^2)).
\label{eqn:forcej}
\end{align}
The presented model allows us to efficiently calculate the force onto $N$ beads for several different parameters. However, a full analytic treatment is rather tedious and we will focus on some special cases of reduced complexity before including all possible processes in the following. A concrete generic example can be found in Appendix A. 

\section{Forces for two forward propagating modes}
While for very small point like scatterers there is a symmetry between forward and backward scattered light~\cite{asboth2008optomechanical}, for larger or thicker objects forward scattering is usually dominant with respect to reflection~\cite{sonnleitner2011optical,maimaiti2016nonlinear} and the coupling between modes of the same propagation direction dominates (see e.g. Appendix A). Therefore we first consider larger particles of lower contrast and restrict our treatment to two different transverse forward propagating modes and injection from a single side. This already will show the essential physics without requiring a too complex analytic form. 

\subsection{Single particle}
To study the fundamental effects of  transverse forward mode coupling, we first investigate only a single particle in the fibre field, which will introduce coherent coupling between the two propagating modes. In this special case the nontrivial part of the four port matrix~\eqref{eqn:M4p} reduces to a two port matrix and the amplitudes left and right of the beam splitter are connected via the following matrix
\begin{align}
\begin{pmatrix}
C_1\\C_2
\end{pmatrix}
=\textbf{M}_{2p}\begin{pmatrix}
A_1\\A_2
\end{pmatrix}=
\begin{pmatrix}
t_{11} e^{i\phi_{11}} & t_{21} e^{i\phi_{21}}\\
t_{12} e^{i\phi_{12}} & t_{22} e^{i\phi_{22}}\end{pmatrix}\begin{pmatrix}
A_1\\A_2
\end{pmatrix},
\end{align} 
where we chose $A_1\equiv A_{1,1}$, $A_2\equiv A_{2,1}$, $C_1\equiv A_{1,1}$ and $C_2\equiv C_{2,1}$. The scattering process changes the relative mode amplitudes and adds a light phase via $e^{i\phi_{ij}}$. Here we can choose $\phi_{11}=0$ without any restriction and for a single particle also $\phi_{22}$ will not change the force expression. 

The unitarity condition $\mathbf{M}_{2p}^\dagger \cdot\mathbf{M}_{2p}=\mathbb{I}$ then reduces the number of physically relevant parameters to
\begin{align}
t_{11}&=t_{22}=t,\\
t_{12}&=t_{21}=\sqrt{1-t^2},\label{eqn:t12fromt}\\
\phi_{12}&=-\phi_{21}+(2n+1)\pi=\phi,~n\in\mathbb{N},
\end{align}
which finally leads to a much simpler form of the scattering matrix
\begin{equation}
\textbf{M}'_{2p}=\begin{pmatrix} t &  -e^{-i\phi}\sqrt{1-t^2} \\ e^{i\phi}\sqrt{1-t^2} & t \end{pmatrix}.
\label{msc}
\end{equation}
As a result the output amplitudes are connected with the input amplitudes via
\begin{align}
C_{1} &= t A_{1}+e^{i\phi}\sqrt{1-t^2} A_{2},\\
C_{2} &= -e^{-i\phi}\sqrt{1-t^2} A_{1}+ t A_{2}.
\label{C12}
\end{align}

To calculate the force we insert this result into~\eqref{eqn:forcej}  which leads to the force onto a single particle in two forward propagating modes
\begin{align}
F_\mathrm{SP}^{2p}=\frac{cn\epsilon_0\hbar}{2}&(k_2-k_1) \left(\left(t^2-1\right)\left(\vert A_1\vert^2-\vert A_2\vert^2\right)\right.\nonumber \\
&\left.+t\sqrt{1-t^2}\left(e^{i\phi}A_1^*A_2+e^{-i\phi}A_1A_2^*\right)\right).
\label{eqn:F1}
\end{align}
The two wavenumbers ($k_1$ for the fundamental and $k_2$ for the higher order mode) differ in general and fulfil the condition $k_1>k_2$.

\begin{figure}
\centering
\includegraphics[width=0.45\textwidth]{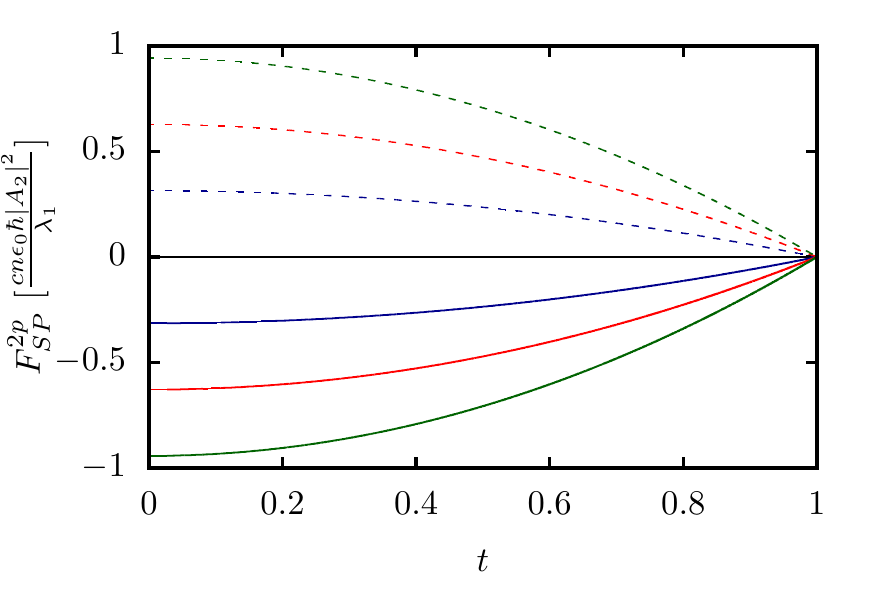}
\caption{Force on a single particle in only forward propagating modes considering only the fundamental (dashed line) or the higher order mode field (solid line) as input. The blue line corresponds to $k_2=0.9~k_1$, the red line to $k_2=0.8~k_1$ and the green line to $k_2=0.7~k_1$.}
\label{F2porthom}
\end{figure}

Let us now look at the special translation invariant case, where only one of the two modes is pumped and the other mode is only populated by scattered photons. There are two possible cases to consider. On the one hand only the fundamental mode can be pumped ($A_1\neq0$, $A_2=0$) and on the other hand only the higher order mode can be injected ($A_1=0$, $A_2\neq0$). The corresponding forces are shown in~\fref{F2porthom}. One finds that the force on the particle is always negative, if the higher order mode is chosen as the input field. The negative force implies that for this configuration the particle is pulled in the direction against the incoming beam. It is fundamental to state that due to translation symmetry of the system, this force is the same anywhere along the fibre and the particle will be continuously pulled towards the source over very long distances. 

Mathematically this behavior gets obvious, if one looks at the analytic expression for the force~\eqref{F1hom}. Assuming only a higher order mode input field ($A_1=0$, $A_2\neq0$) the force simplifies to
\begin{align}
F_\mathrm{SP}^{2p}|_{A_1=0}=\frac{cn\epsilon_0\hbar}{2}(k_1-k_2)\left(t^2-1\right)\vert A_2\vert^2.
\label{F1hom}
\end{align}
As mentioned above the conditions $0\le t\le 1$ and $k_2<k_1$ have to hold. Consequently, the force is always negative in this case. Following the same procedure but using the fundamental mode field as the input beam, leads to a positive force as depicted in~\fref{F2porthom} following the conventional expectation.

Lets now discuss the essential physics leading to this somewhat counterintuitive result. Due to the fact that the higher order mode contains a higher amount of transverse momentum and thus less longitudinal momentum than the fundamental mode, the scattering process between those modes (with amplitude $t_{12}$) is generally suppressed. Only some additional momentum provided by the bead closes this momentum gap and allows for scattering between the two modes. This is quite analogous to the case of Bessel tractor beams~\cite{ruffner2012optical}, but due to the presence of a light confining geometry the range of the effect is infinite in the studied case.

\subsection{Two particles}
As it has been seen and experimentally demonstrated in previous work~\cite{maimaiti2016nonlinear} the two mode forward scattering model also well describes the long range inter-particle forces in this system. In the simplified case with no backscattering, however, the fields and the force on the first particle cannot be affected by a second one downstream the mode. Nevertheless, the field impinging on the second particle is the result of the mode mixing by the first and thus significantly depends on the relative distance allowing for stable configurations. Using the scattering matrix approach the outgoing amplitudes after the second particle can be calculated in the following manner:
\begin{equation}
  \begin{pmatrix} C_{1,2}\\C_{2,2}\end{pmatrix}= \textbf{M}_{2p}'\cdot\textbf{P}(d)\cdot\textbf{M}_{2p}'\begin{pmatrix}  A_{1,1}\\A_{2,1}\end{pmatrix},
\end{equation}
with $A_{i,j}$ and $C_{i,j}$ being the input and output fields with mode $i$ at the $j$-th particle. In addition, the free field propagation over a distance $d$ (\ie the propagation face shift) is included via a transfer matrix which connects the amplitudes on the right hand side of the first and the left side of the second beam-splitter
\begin{equation}
\textbf{P}(d)=\begin{pmatrix}
e^{ik_1d}&&0\\0&&e^{ik_2d}.
\end{pmatrix}
\end{equation}
Here we assume identical particles for simplicity so that the scattering process is the same for both particles. Of course, the presented scheme can be generalized to more particles in a straightforward manner. The force on the $j$-th particle is then again defined by~\eref{eqn:forcej}.

\begin{figure}
\centering
\includegraphics[width=0.45\textwidth]{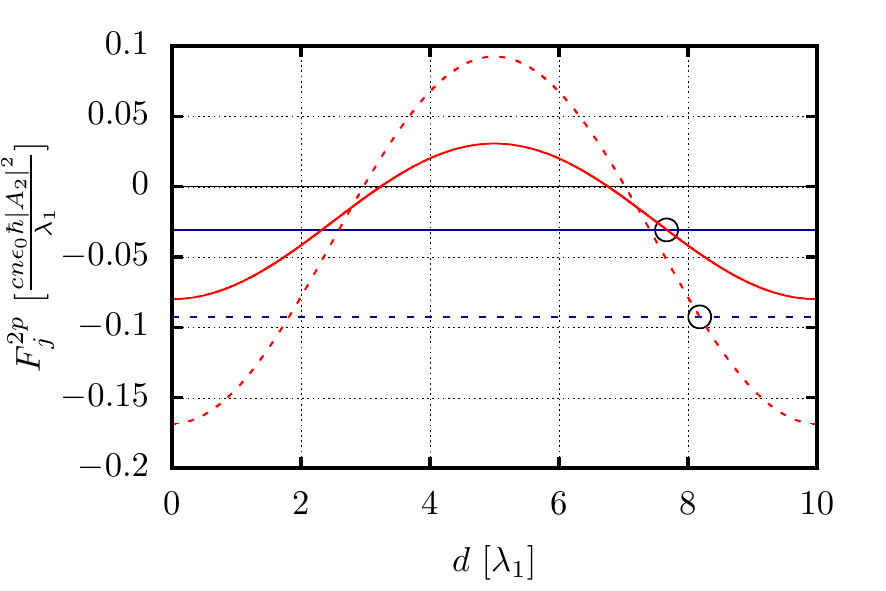}
\caption{Force on the first (blue) and second particle (red) in only forward propagating modes considering only the higher order mode field as input and choosing $k_2=0.9~k_1$. The solid line corresponds to $t=0.95$, thus~\eref{eqn:t12fromt} implies $t_{12}=0.31$. The dashed line shows the results for $t=0.84$ and $t_{12}=0.54$. The black circles indicate stable distances.}
\label{F2port2part}
\end{figure}

To keep the discussion compact, we will restrict our treatment here to the two particle case. Due to the fact that the analytical expression of the forces is rather lengthy, it will be omitted here but the spatial dependence of the forces is presented in~\fref{F2port2part}. Interestingly, one finds that for certain distances between the two particles, both particles can experience a negative or tractor force. Obviously as a next step one can ask for stable configurations of the two particles,~\ie configurations in which the particle's distance $d$ remains constant and locked against small perturbations. This requires both forces to have the same value ($F_1-F_2=0$) and at the same time one needs $\frac{\partial F_1}{\partial d}>0$ and $\frac{\partial F_2}{\partial d}<0$,  i.e. a restoring force if the particles deviate from this equilibrium equal force position.

As an example we again choose the higher order mode as the input field ($A_1=0$, $A_2\neq0$). Interestingly, here stable distance configurations with a negative force on both particles can be found for the chosen parameters. Therefore, the particles are commonly attracted towards the beam source while they stay at a constant distance. This corresponds to a stable collective tractor beam configuration for the particles. Another interesting fact, which can be seen in~\fref{F2port2part} is that for certain distances beyond the stable point, the tractor force on the second particle can even be stronger than the one on a single particle. Hence the mode phase lock introduced by the first particle creates an even stronger trap than on itself.
\begin{figure}
\centering
\includegraphics[width=0.4\textwidth]{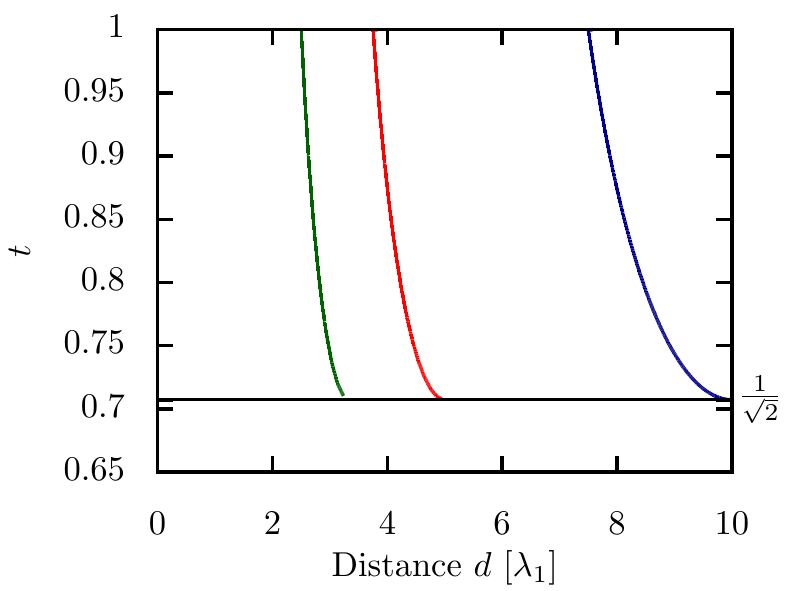}
\caption{Optical binding distances $d$ as a function of the transmission coefficient $t$ for two particles in forward propagating modes considering only the higher order mode field as input. The blue line corresponds to $k_2=0.9~k_1$, the red line to $k_2=0.8~k_1$ and the green line to $k_2=0.7~k_1$.}
\label{F2port2partc}
\end{figure}

Note the lower cutoff of the distances in ~\fref{F2port2partc}, which reflects the fact that stable tractor beam configurations cannot be found for values of $t$ below certain values. In particular the transmission $t$ by the first particle has to exceed a certain threshold value in order to find a stationary two particle configuration. This is due to the fact that if too much light is scattered to the fundamental mode by the first particle, the relative mode amplitude ratio in the fiber between the two particles will be rather small and the tractor force on the second particle will always be too small to follow the first one against the beam propagation direction at a constant distance. Note that the particle force can also be seen as the simplest possible form of optical binding as the scattering of the first particle creates a series of dipole traps for the second one.   

\section{Four mode model including backscattering}

Lets now turn to a more realistic model. While the relative magnitude of backward versus forward scattering can be small for large particles at low index contrast, it will of course never be exactly zero~\cite{maimaiti2016nonlinear}. As the momentum transfer per photon in reflection is much larger than for forward mode mixing even a small amplitude could induce some changes. In this section we will thus treat the whole 4-mode model as already described above and investigate, how much even small backscattering at the particles position can induce substantial force contributions. The scattering processes in this case are described by the general 4-port matrix~\eqref{eqn:M4p}. Losses to other modes or absorption also can be included by modifying the diagonal of this matrix.

Obviously the set of available parameters is rather large in this case and hardly can be exhaustively treated. Nevertheless, at least in principle for any concrete bead size, position,  shape and mode geometry, they could be at least numerically calculated from a generalized Helmholtz equation. Here we will use a different approach and look at physically interesting parameter ranges. Once interesting parameters are found, one could look for geometries, where one could implement such scattering properties. In order to reduce the complexity we will first assume that both modes experience the same forward scattering amplitudes and phase shifts with negligible backscattering into the very same mode. This  allows us to concentrate on the effects of mode cross-scattering processes which are at the origin of interesting multi-mode physics. A concrete example is presented in the appendix.

Based on the above arguments we first set: 
\begin{align}
t_{11}&=t_{22}=t,\\
r_{11}&=r_{22}=0.
\end{align}
We again also set the reference phases to zero ($\phi_{11}=\phi_{22}=0$). The unitary condition for the full coupling matrix then again gives the following set of necessary conditions
\begin{align}
t_{12}&=t_{21},\\
r_{12}&=r_{21},\\
t_{12}&=\sqrt{1-t^2-r_{12}^2}, \label{eqn:t12fromtr}\\
\phi_{21}&=-\phi+(2m-1)\pi,\\
\psi_{21}&=-\phi-(n+1/2)\pi,\\
\psi_{12}&=\phi+(n-1/2)\pi,~~ m,n\in\mathbb{N},
\end{align}
where we defined $\phi=\phi_{12}$. In this case the scattering matrix simplifies to
\begin{equation}
\textbf{M}_{4p}'=\begin{pmatrix}
 t & -e^{-i \phi } t_{12} & 0 & i e^{-i \phi } r_{12} \\
 e^{i \phi } t_{12} & t & i e^{i \phi } r_{12} & 0 \\
 0 & i e^{-i \phi } r_{12} & t & -e^{-i \phi } t_{12} \\
 i e^{i \phi } r_{12} & 0 & e^{i \phi } t_{12} & t
\end{pmatrix}.
\label{eqn:M4psimple}
\end{equation}
Of course the two-port system as it has been investigated in the previous section can be reproduced by setting $r_{12}=0$.

\subsection{Single particle}
We again start with investigating the forces acting on a single particle along the fiber. We follow the same procedure as presented above but with more amplitudes coupled by a larger matrix. Using~\eqref{eqn:forcej} in order to calculate the force leads to
\begin{align}
F_\mathrm{SP}^{4p}=&\frac{cn\epsilon_0\hbar}{2}\left(\vert A_1\vert^2 \left(k_1\left(r_{12}^2+t_{12}^2\right)+k_2 \left(r_{12}^2-t_{12}^2\right)\right)\right.\nonumber\\
&\left.+\vert A_2\vert^2 \left(k_1 \left(r_{12}^2-t_{12}^2\right)+k_2\left(r_{12}^2+t_{12}^2\right)\right)\right.\nonumber\\
&\left.+t_{12}t \left(k_1-k_2\right)\left( A_1A_2^*e^{i \phi }+ A_1^*A_2 e^{-i \phi }\right)\right).
\end{align}

\begin{figure}
\centering
\includegraphics[width=0.45\textwidth]{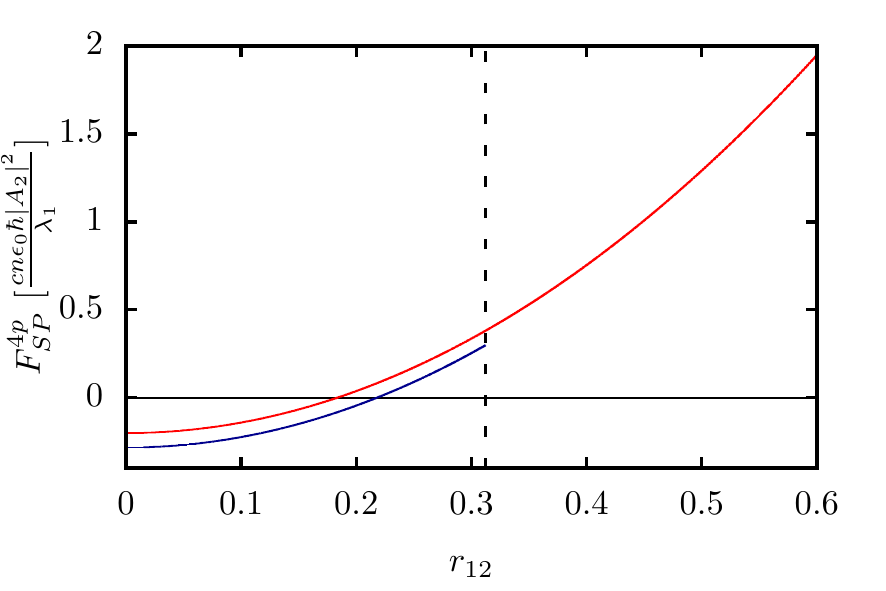}
\caption{Force on a single particle including back-reflection and considering only the higher order mode field as input field and $k_2=0.9~k_1$. The blue line corresponds to $t_{12}=0.95$ and the red line to $t_{12}=0.8$. The dashed line illustrates the threshold on the reflection $r_{12}$ defined by~\eref{eqn:t12fromtr}.}
\label{F4port}
\end{figure}

For injection of only a higher order mode field ($A_1=0$) this reduces to
\begin{align}
F_\mathrm{SP}^{4p}  = \frac{cn\epsilon_0\hbar}{2}\left[t_{12}^2\left( k_2-k_1\right)+r_{12}^2 \left(k_1+k_2\right)\right]\vert A_2\vert^2,
\end{align}
which can be negative as long as the parameters fulfil the following condition:
\begin{align}
\frac{r_{12}^2}{t_{12}^2} \le \frac{k_1-k_2}{k_1+k_2}.
\label{eqn:rthres}
\end{align}

Fig.~\ref{F4port} shows an example how the backscattering process gives a negative force on the particle as long as $r_{12}$ is not too large. Note that as the two momenta are usually not too different the condition on the smallness of back-reflection can be rather stringent.

\subsection{Two particles}

As the scattering matrix~\eqref{eqn:M4psimple} expresses the outgoing fields in terms of the incoming fields the expression for the fields generated by two beads formally looks similar to the case above. However, to find the amplitudes on two beads, the total transfer matrix $M_{TM}$ has to be found in accordance with the prescribed boundary conditions. Here one immediately gets an infinite series of reflections and back reflections. As shown in previous work~\cite{ostermann2014scattering} this can be efficiently calculated by rearranging the terms of the matrix in a form which connects the amplitudes to the left and to the right of a particle in the form 
\begin{equation}
\textbf{M}_{TF}=\begin{pmatrix}
      \frac{t}{1-r^2_{12}} & \frac{i r_{12}t_{12}}{1-r^2_{12}} & -\frac{t_{12}e^{- i \phi}}{1-r^2_{12}} & -\frac{ir_{12}t e^{-i \phi}}{1-r^2_{12}} \\\frac{i r_{12}t_{12}}{1-r^2_{12}} &\frac{t}{1-r^2_{12}} &\frac{ir_{12}t e^{-i \phi}}{1-r^2_{12}} &\frac{t_{12}e^{- i \phi}}{1-r^2_{12}} \\ \frac{t_{12}e^{i \phi}}{1-r^2_{12}}& -\frac{i e^{i\phi}r_{12}t }{1-r^2_{12}} &\frac{t}{1-r^2_{12}} &-\frac{i r_{12}t_{12}}{1-r^2_{12}}\\ \frac{ir_{12}t e^{i \phi}}{1-r^2_{12}} &-\frac{t_{12}e^{ i \phi}}{1-r^2_{12}} & -\frac{i r_{12}t_{12}}{1-r^2_{12}}& \frac{t}{1-r^2_{12}}
      \end{pmatrix} ,
\end{equation}
so that
\begin{equation}
 \begin{pmatrix} C_{1,2}\\D_{1,2}\\C_{2,2}\\D_{2,2}\end{pmatrix}=
  \textbf{M}_{TF}\cdot\textbf{P}_4(d)\cdot\textbf{M}_{TF}
  \begin{pmatrix}
  A_{1,1}\\B_{1,1}\\A_{2,1}\\B_{2,1}\end{pmatrix},
\end{equation}
where $\mathbf{P}_4(d)$ is the 4 mode generalized propagation matrix for the fields:
\begin{equation}
\textbf{P}_{4}(d)=\begin{pmatrix}
      e^{ik_1d} & 0 & 0 & 0 \\0 &e^{-ik_1d} &0 &0 \\ 0& 0 &e^{ik_2d} &0\\0 &0 & 0& e^{-ik_2d}
      \end{pmatrix}.
\end{equation} 
In this rearranged form, the amplitudes on the particles can be found by solving these four equations for the required input and output fields.
\begin{figure}
\centering
\includegraphics[width=0.45\textwidth]{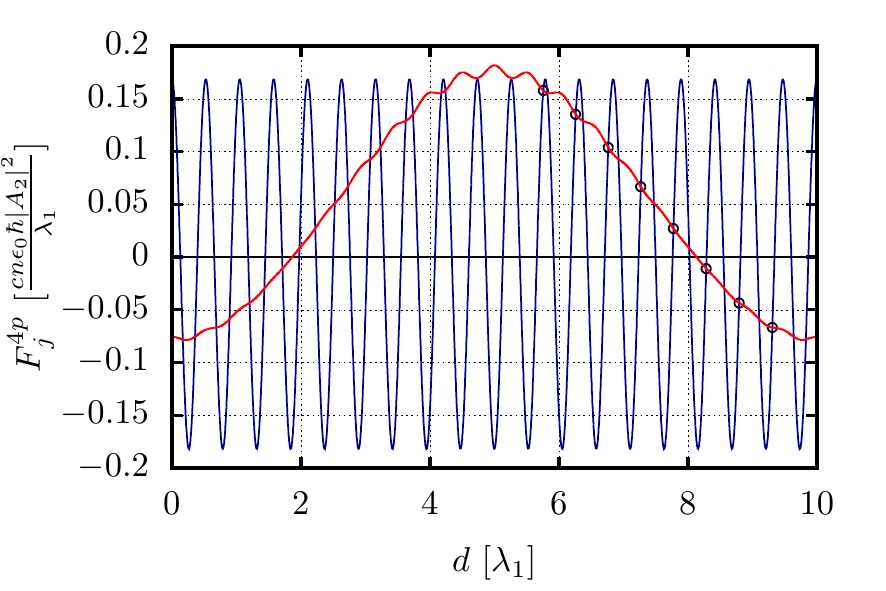}
\caption{Force on two particles including back-reflection and considering only the higher order mode field as input field and $k_2=0.9~k_1$. The blue line corresponds to the force on the first and the red line to the force on the second particle. Here, we chose $t_{12}=0.54$ and $r_{12}=0.12$. Black circles indicate stable points.}
\label{F4port2part}
\end{figure}
As the analytical expression even for the single particle force is rather complex, we will not give it here and simply show some numerical examples. We see that the presence of the second particle changes the interaction between the particles and the fields in a much more complex way as above. Indeed, both particles now are influenced by the presence and position of the other one. Indeed, as demonstrated in Fig.~\ref{F4port2part} the force on the second particle now exhibits small but spatially fast fluctuations due to the interference of the reflected fields in addition to the larger oscillations from mode beating. Also the force on the first particle is oscillating now. These oscillations are a consequence of the interference between the incoming and backscattered fields. Here we see that the effective total backscattered field can be less than for a single bead as the field amplitudes from the two particles interfere like in a Fabry-Perot resonator driven with a resonant wavelength. Hence we can find parameter ranges with an even bigger tractor force than for a single particle. This can be enhanced by a collective mode coupling of the two particles. 
 
\begin{figure}
\centering
\includegraphics[width=0.43\textwidth]{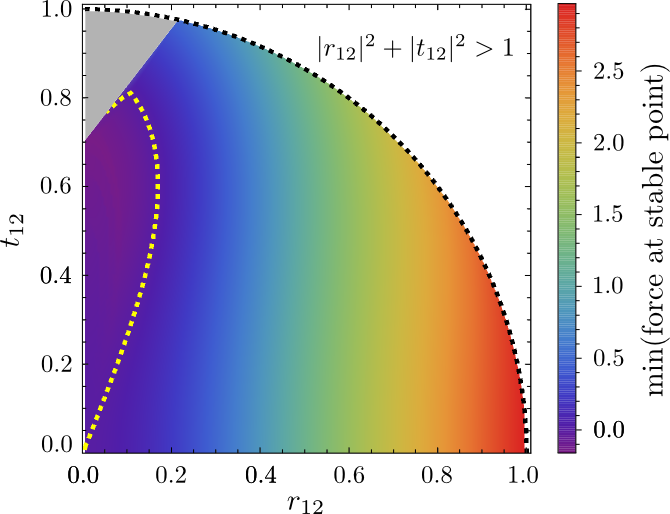}
\caption{The minimal possible force at a stable point plotted for different values of $r_{12}$ and $t_{12}$. The yellow dashed line marks the zero line. In the grey region no stable solution exists at all and the black dashed line shows the threshold for reflection and transmission imposed by~\eref{eqn:t12fromtr}.}
\label{fig:minforcestable}
\end{figure}

In general the distance of maximal tractor force on the first particle does not correspond to a stable distance, but we find a large number of potentially stable distances. A fair fraction of these correspond to a net negative force on both particles. Hence, we see that also the full model allows for a many particle tractor beam configurations.

If we analyze this fact in more detail, we find that there exists a certain region of parameters in which a stable two particle tractor beam, i.e. a collectively enhanced tractor, can be realized. This region corresponds to the one on the left side of the yellow dashed line in~\fref{fig:minforcestable}. Obviously, in order to establish a negative tractor beam force the reflection as well as the transmission coefficients between the two modes must not exceed a certain critical value. On the one hand if the transmission is above a certain critical value no more stable point (\ie points which fulfil the condition $F_1-F_2=0$) can be found.) On the other hand, if the reflection coefficient exceeds a certain value the force can no longer be negative and the particles will always be pushed in the direction of the incoming laser beam. This effect is related to the result for two particles in forward propagating fields (\cf~\eref{eqn:rthres}). Nevertheless, due to the fact that the back-reflection for extended objects is in general rather small, the necessary parameter regime can be reached in some specific geometries like for example for beads trapped inside a two-mode hollow core fibre (see Appendix A).

\section{Conclusions and outlook}

We demonstrated that optical fibers supporting at least two transverse modes can be the basis of translation invariant tractor beam implementations dragging particles and even pairs of optically bound particles against the injected beam direction without the need of any external control or feedback. The results are easily generalizable to several particles. In practical implementations the remaining backscattering seems to be one of the central obstacles to overcome here. For the implementation, only the higher order mode has to be pumped via the one end of the fibre, where the particles should be transported with hardly any restrictions on the bandwidth or coherence length. Apart from being a neat physical mechanism, the tractor effect could be helpful in setups where one plans to extract particles from a trap source and load them into a dipole trap at the other end of the fiber\cite{ vorrath2010efficient}.

In this work we have primarily shown that such tractor beams are theoretically possible and given the necessary boundary conditions on the parameters to be achieved. The main challenge in practice actually is to design the fibre in a way to maximize cross coupling between two modes and minimizing loss to others. In a multi-mode fibre the tractor force will be the bigger the higher the order of the injected mode and the larger the amplitude of the fundamental mode at the particle position. For a hollow core fibre coupling should be good using modes of the same symmetry as the $TEM_{00}$ and $TEM_{20}$ mode with a particle of about wavelength size at its center as shown in the appendix.

\acknowledgments
We thank C. Genes  for helpful discussions and acknowledge support by the Austrian Science Fund FWF through projects SFB FoQuS P13 and I1697-N27. We also acknowledge fruitful discussions from  S. Nic Chormaic,  V. G. Truong and A. Maimaiti from the Okinawa Institute of Science and Technology Japan.

%

\appendix
\section{}
A small particle placed inside the mode field of an optical waveguide or optical fiber will locally perturb the field and change its propagation. If one considers positions significantly left and right of the bead, where near field effects can be neglected, the propagation field can still be expanded in terms of the transverse eigenmodes of the waveguide. For a bead made of linear optical material the corresponding expansion coefficients on both sides are linearly connected by an effective transfer matrix. 

The effect of the particle will be phase and amplitude changes of the modes induced by light scattering between various forward and backward propagation modes. In addition, some of the field could be absorbed or, equivalently, scattered to free space modes not guided by the fiber so that the transfer matrix is not necessarily unitary. As the latter effect cannot positively contribute to the tractor effect or inter-particle forces we will neglect it for this calculation of the mode coupling coefficients.

To obtain the coefficients one needs to solve the corresponding Helmholtz equation including proper boundary conditions, which for a detailed modelling requires rather extensive numerical calculations as e.g. presented in~\cite{maimaiti2015higher,maimaiti2016nonlinear}. Here we are mainly interested in the basic physics of the tractor force effect, which is contained in the relative magnitude of the effective coupling between different transverse modes. As shown above a strong tractor force requires stronger coupling of a mode to lower order modes than to higher order modes as well as very low reflection. For only two modes the coupling between the modes needs to be stronger than free space losses.

\begin{figure}
\centering
\includegraphics[width=0.45\textwidth]{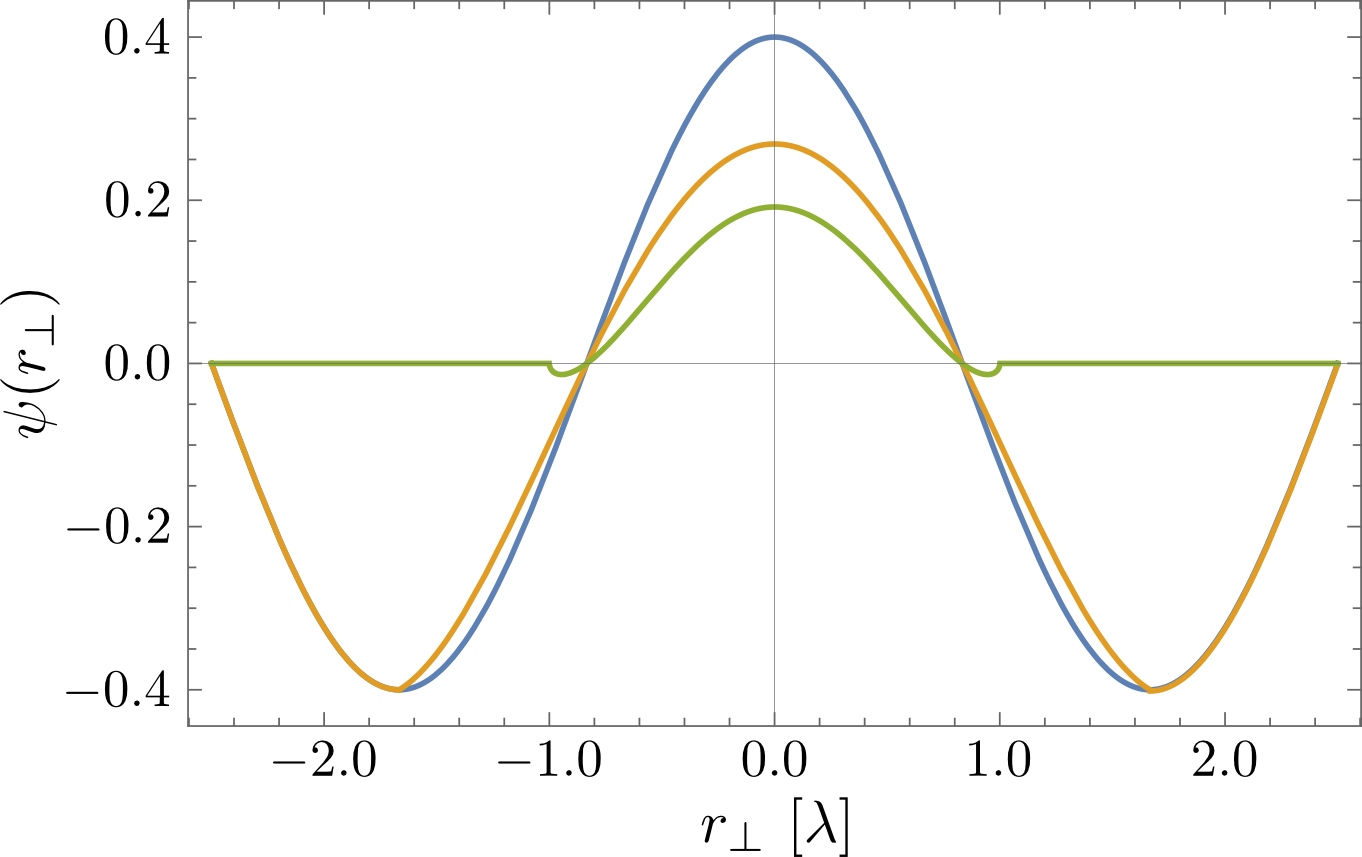}
\caption{Shape of the original (blue) and distorted mode function (real part orange, imaginary part green) for a spherical bead of refractive index $n=1.25$ and diameter $2 \lambda$ in the center of a rectangular waveguide for a waveguide diameter  $a=5 \lambda$.}
\label{fig:distorted}
\end{figure}
      
In order to keep the complexity low we thus refrain to a paraxial approximation for the propagation of the fields. In this case the field evolution along a non-absorbing bead can be approximated by the effective optical Schr\"odinger equation, where the bead creates an effective potential for the transverse light field  amplitude $\psi(r_{\perp}) $ propagating along the fiber direction z~\cite{lamprecht1998comparing}: 
\begin{equation}
\frac{i}{n_0 k_0}\partial_z \psi(r_{\perp}) =\left[\frac{-\Delta_{\perp}}{2 n_0 k_0^2} + V_{opt}(r_{\perp})\right] \psi(r_{\perp}).
\end{equation}
In this case the optical potential of the bead 
\begin{equation}
V_{opt}(r_{\perp}) =  \frac{n_0^2-n(r_{\perp}^2)}{ 2 n_0^2}
\end{equation}
creates an effective local attractive potential which couples the various transverse modes. Here $n(r_{\perp}))$ is the bead refractive index distribution and $n_0$ the background refractive index.

Below we will try to identify a favorable case for generating a tractor force in a simple configuration. To show the qualitative behaviour we simply assume a perfect square waveguide with with a certain diamater $a$~\cite{horak2003giant} or a hollow guide with sharp (metallic) boundaries~\cite{gomez2001resonant} where the mode functions are simply given by harmonic functions vanishing at the boundaries.

\begin{figure}
\centering
\includegraphics[width=0.45\textwidth]{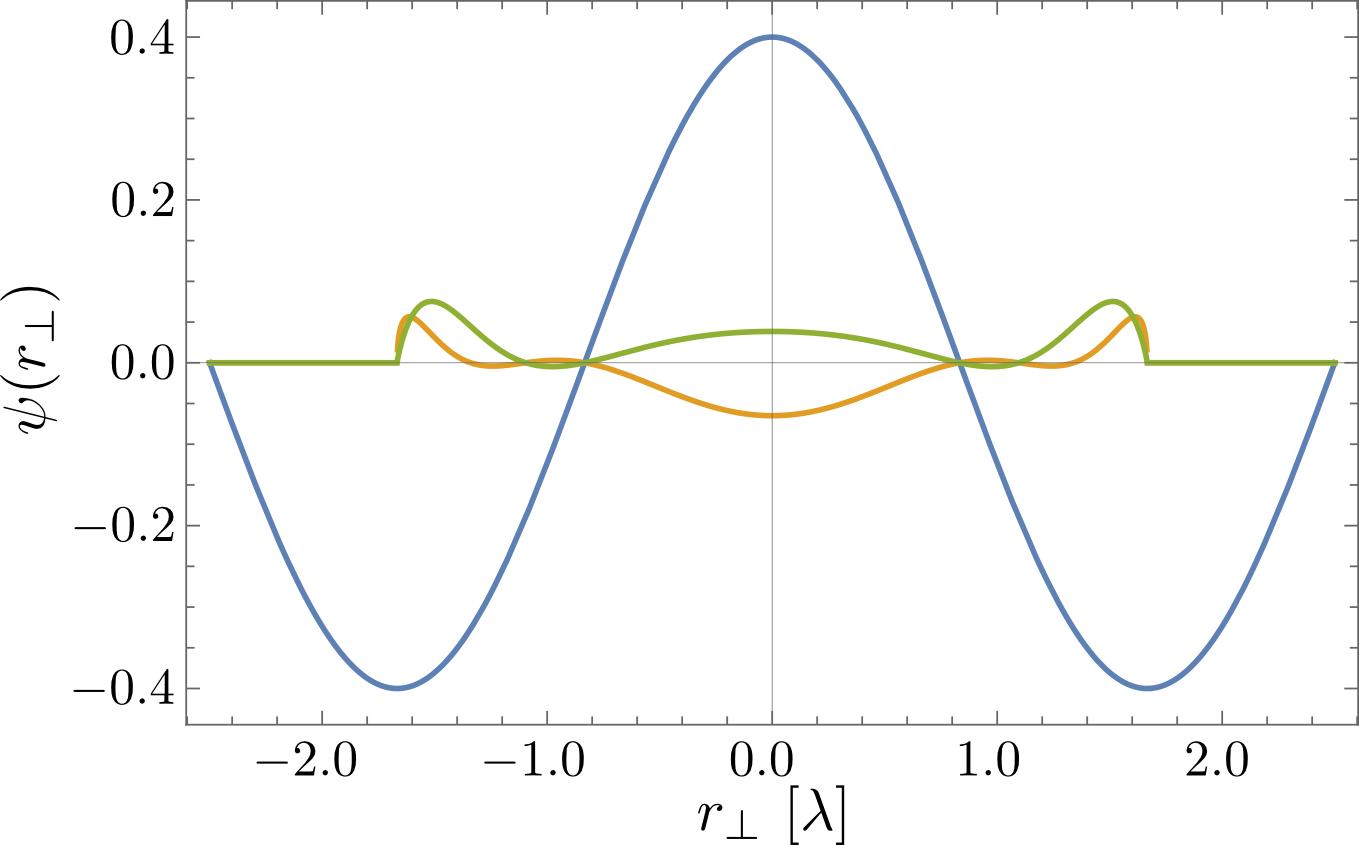}
\caption{Shape of the original (blue) and the reflected mode function for a spherical bead of refractive  index $n=1.25$ and diameter $2 \lambda$ in the center of a rectangular waveguide for a waveguide diameter $a=5 \lambda$.}
\label{fig:reflected}
\end{figure}

For a small refractive index of the bead and a not too large size, the effect of a bead of diameter $d$ on the field can then be simply estimated by the spatially accumulated phase shift  
\begin{equation}
e^{i \phi(r_{\perp})} \approx e^{ i \int_0^d V_{opt}(r_{\perp} ) dz}
\end{equation}
by field while traversing the bead and a small reflected component~\cite{domokos2002quantum}. This is shown in the example in~\fref{fig:distorted} below, where we plot the original third order mode function and its distorted form (real and imaginary part) after the bead on a cut along the x-axis. Similarly we can estimate the reflected contribution as shown in~\fref{fig:reflected}.
\begin{figure}
\centering
\includegraphics[width=0.45\textwidth]{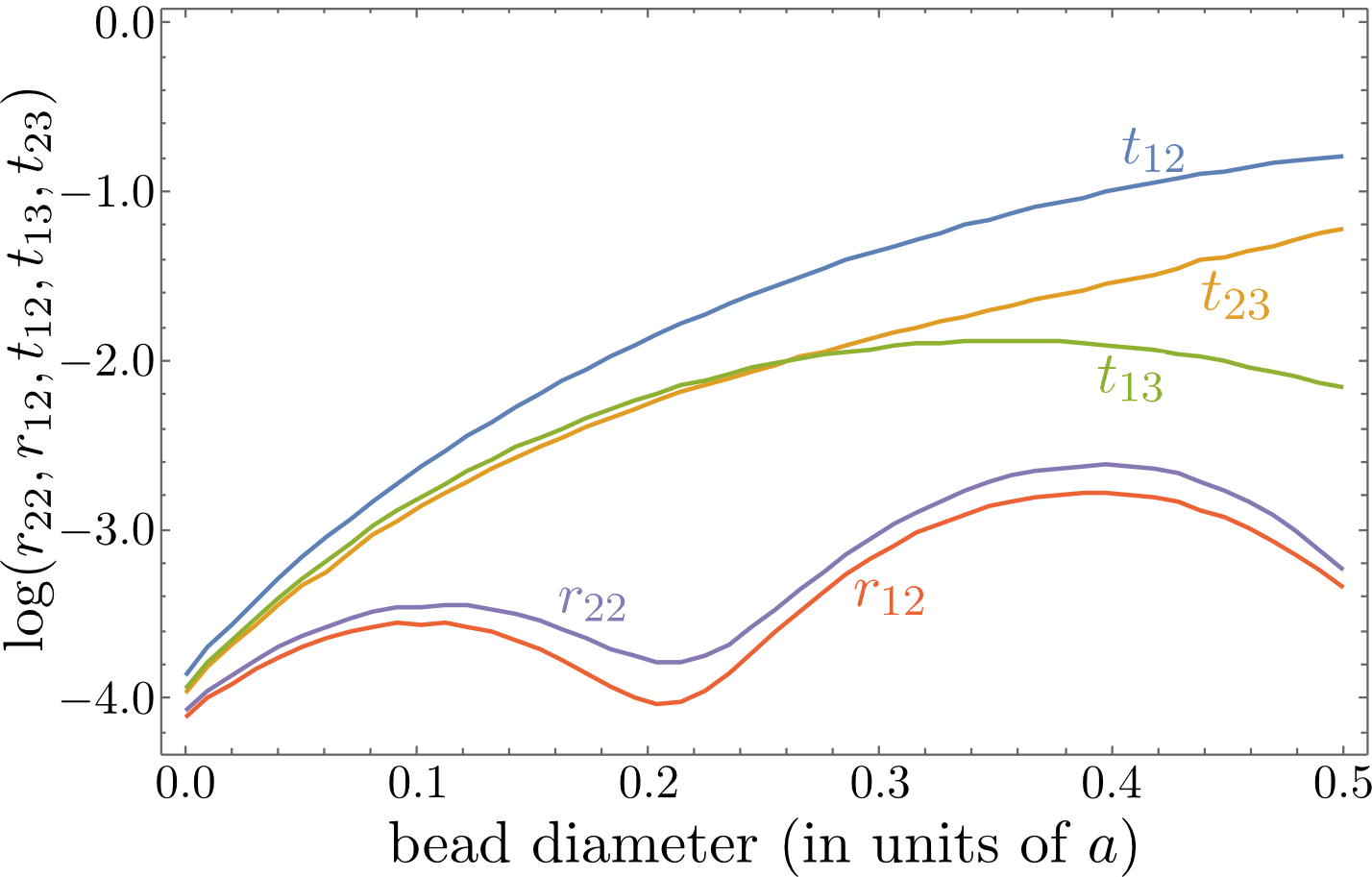}
\caption{Mode coupling coefficients for the three lowest order symmetric modes for a spherical bead of index $n=1.5$ in the center of a rectangular waveguide as function of bead diameter in units of the waveguide diameter $a=9 \lambda$.}
\label{fig:modecouple}
\end{figure}
In this limit the mode coupling coefficients can then be simply obtained by projecting the distorted and reflected fields onto the original modes. This will strongly depend on position, size and refractive index of the bead. If we want strong overlap between a higher order and a lower order mode it is thus favourable to put the bead at a position, where the target mode has a high amplitude but the unwanted modes are low or strongly varying. As an example here in~\fref{fig:modecouple} we show the case of a bead of varying size exactly at the center of the waveguide where the first and third order mode amplitudes are large, while others are small. We see that indeed the desired coupling between mode one and two is about one order of magnitude bigger than the  other couplings and reflections. Hence, this configuration should lead to a sizeable tractor force for a suitable particle size range.

\end{document}